\documentclass[useAMS,usenatbib,usegraphicx]{mn2e}

\title[Spectral Index of IRAS 17347-3139]{Spectral Index of the H$_2$O-Maser
  Emitting Planetary
  Nebula IRAS 17347-3139}
\author[J. F. G\'omez et al.]{J. F. G\'omez,$^{1,2}$ I. de
  Gregorio-Monsalvo,$^{2}$ J. E. J. Lovell,$^3$
  G. Anglada,$^1$
\newauthor
L. F. Miranda,$^1$ O. Su\'arez,$^{2}$ J. M. Torrelles,$^{4,5}$
Y. G\'omez$^6$\\
$^{1}$Instituto de Astrof\'{\i}sica de Andaluc\'{\i}a, CSIC, Apartado
3004, E-18080 Granada, Spain\\
$^{2}$Laboratorio de Astrof\'{\i}sica Espacial y F\'{\i}sica Fundamental,
  INTA, Apartado 50727, E-28080 Madrid, Spain\\
$^{3}$Australia Telescope National Facility, CSIRO, P.O. Box 76,
Epping, NSW 1710, Australia\\
$^{4}$Instituto de Ciencias del Espacio (CSIC)-IEEC, Gran Capit\`a
2-4, 08034 Barcelona, Spain.\\
$^{5}$On
  sabbatical leave at the UK Astronomy Technology Centre, Royal
  Observatory Edinburgh\\
$^{6}$Centro de Radioastronom\'{\i}a y Astrof\'{\i}sica, UNAM, Apartado Postal
3-72 (Xangari), 58089 Morelia, Michoac\'an, Mexico}
\begin{document}

\date{\today}


\maketitle

\begin{abstract}
We present radio continuum observations of the planetary nebula (PN) IRAS
17347-3139 (one of the only two known to harbour water maser emission), made
to derive its spectral index and the turnover frequency of the
emission. The spectrum of the source rises in the whole frequency range sampled, from 2.4 to 24.9 GHz, although the spectral index
seems to decrease at the highest
frequencies ($0.79\pm 0.04$ between 4.3 and 8.9 GHz, and $0.64\pm
0.06$ between 16.1 and 24.9 GHz). This suggests a turnover frequency
around 20 GHz (which is unusual among PNe, whose radio emission usually
becomes optically thin at frequencies $< 10$ GHz), and a relatively
high emission measure ($1.5\times 10^9$ cm$^{-6}$ pc). The radio continuum
emission has increased by a factor of $\simeq 1.26$ at
8.4 GHz in 13 years,  
which can be explained as
expansion of the ionized region by a factor of $\simeq 1.12$ in radius with a
dynamical age of $\simeq 120$ yr and at an expansion 
velocity of $\simeq 5-40$ km s$^{-1}$.
These radio continuum characteristics, together with
the presence of water maser emission and a
strong optical extinction suggest that IRAS
17347-3139 is one of the youngest PNe known, with a relatively massive
progenitor star. 
\end{abstract}


\section{Introduction}

Maser emission of different molecules (SiO, OH, H$_2$O) is observed
in evolved stars, from red giants through to planetary nebulae
(PNe). This emission seems to be stratified in the envelopes of AGB
stars, with SiO masers
located close to the stellar surface, water masers at $\simeq 10-100$
AU, and OH masers further away, up to $\simeq 10^4$ AU \citep{rei81,lan87}. In the
evolution of an AGB star to a PN, masers disappear
sequentially, starting from the innermost ones \citep{lew89}. Only water and OH
masers are generally found in proto-PNe, but as the star enters its PN phase,
water molecules are rapidly destroyed by the ionizing
radiation \citep{lew89,gom90}. Therefore water masers are not
expected in PNe unless these are extremely young.
The first detection of water maser emission in a PN was obtained 
in K3-35 \citep{mir01}. Maser emission in this source is observed
distributed in a
disk-like structure around the central star as well as at the tips of
a bipolar jet.

Recently \citet[hereafter DG04]{dg04}
carried out a survey for water maser emission toward a
sample of 26 PNe, to check whether water
masers are relatively common toward these kind of objects or if K3-35 is an
extraordinary case. As a result of
this work they detected a new cluster of water masers toward 
the southern source IRAS
17347-3139, which constitutes the second bona-fide PN known to show water
maser emission. The low rate of detection (1 out of 26) is
statistically compatible with the short ($\simeq 100$ yr) lifetime of
water molecules in the envelopes of PNe.

Very Large Array (VLA) maps of IRAS 17347-3139 (DG04) 
show 13 maser components, distributed
in a ellipse with 
axes $\simeq 0\farcs25\times 0\farcs12$ ($\simeq 200\times 100$ AU at 0.8 kpc), with
the 22 GHz radio continuum emission located at one of the tips of its
major axis (DG04). This suggests a  possible binary nature for this
source, with the radio continuum emission
related to the central star, and the water
maser ellipse associated to a companion.

The radio continuum emission of IRAS 17347-3139 shows a spectral index 
$\alpha = 0.76\pm 0.03$ ($S_{\nu} \propto\nu^{\alpha}$) 
between 4.9 and 14.9 GHz (DG04), consistent with a partially optically
thick ionized region. 
However, the flux density at 22 GHz was significantly lower
  than the value expected from the spectral index of 0.76
  found at lower frequencies. This discrepancy could be explained as a
  combination 
  of source variability (the 22 GHz data were taken 11 years later than those
  at lower frequencies), calibration errors (given the low elevation of this
  southern source from the VLA), extended emission ($\ga 5''$) 
not sampled by the  VLA, 
or a change in the opacity regime.

The last possibility is especially
compelling. If the ionized region becomes optically thin between 14.9 and 22
   GHz, one would expect a flat spectral index at frequencies higher than the
   turnover frequency. 
Knowing the turnover frequency of the ionized region would
allow us to measure some of its physical parameters.
Obviously quasi-simultaneous radio continuum observations from a southern
telescope were necessary to rule out variability and calibration
effects, and to
determine whether the turnover frequency of the emission is indeed 
around 20 GHz. The flux density and spectral index of radio-continuum emission can provide
 interesting information about the ionized region and, by inference,
 also about the physical characteristics of the central star. Given
 that water maser emission is a rare phenomenon in PN and that it is
 thought to be present only among the youngest ones it is important
 to study in detail IRAS 17347-3139 to
 see whether we can find some special characteristics that may be
 related to its ability to pump water masers. 

In this paper we present a new, detailed study of the spectral index of IRAS
17347-3139 between 2.4 and 24.9 GHz. In Sec. \ref{sec:obs}  
we describe our observations and we discuss our results in Sec. \ref{sec:dis}.

\section{Observations and results}
\label{sec:obs}

Radio continuum observations of IRAS 17347-3139 were carried out with the
Australia Telescope 
Compact Array (ATCA)\footnote{The Australia  Telescope Compact Array
  is part of the Australia Telescope which is  funded by the
  Commonwealth 
of Australia for operation as a National  Facility managed by CSIRO}
 of the Australia Telescope National Facility, on 2004
March 4 and 10. We observed at 11 different frequencies between
1.4 and 25 GHz on both days. However, the data at 1.4 GHz were not useful due
to interference. The total bandwidth in all cases was 128 MHz, sampled over 14
channels. The source PKS 1934-638 was used as the primary flux
calibrator, while 
IVS B1759-396 was the phase calibrator. Flux densities for those
calibrators are 
shown in 
Table \ref{tab:cals}. The phase center of our observations was located
at $\alpha({\rm J2000})$ = $17^h 38^m 00\fs 586$, $\delta({\rm J2000})$ =
$-31^\circ 40' 55\farcs 67$, the
position of the radio continuum emission from IRAS 17347-3139 (DG04).

\begin{table}
\caption{Flux densities of calibrators}
\label{tab:cals}
\begin{tabular}{llll}
\hline
Frequency & $S_f$$^a$ & $S_{p1}$$^b$ & $S_{p2}$$^c$ \\
(GHz) & (Jy) & (Jy) & (Jy)\\
\hline
2.368 & 11.59 & 1.51 & 1.51\\
4.288 & 6.62 & 1.49 & 1.40 \\
6.080 & 4.40 & 1.56 & 1.44 \\
8.128 & 3.07& 1.62 & 1.55  \\
8.896 & 2.74& 1.63 & 1.57  \\
16.064 & 1.24 & 1.57 & 1.55\\
18.496 & 1.02& 1.56  & 1.53\\
21.056 & 0.88& 1.56 & 1.60 \\
22.976 & 0.80& 1.57 & 1.63  \\
24.896 & 0.74& 1.56 & 1.64   \\
\hline
\end{tabular}

\medskip
$^a$ Assumed flux density of  PKS 1934-638\\
$^b$ Bootstrapped flux density of IVS B1759-396 on 2004 March
4. Uncertainties are $\simeq 0.03$ Jy\\
$^c$ Bootstrapped flux density of IVS B1759-396 on 2004 March 10. Uncertainties are $\simeq 0.03$ Jy
\end{table}

\begin{figure*}
\includegraphics{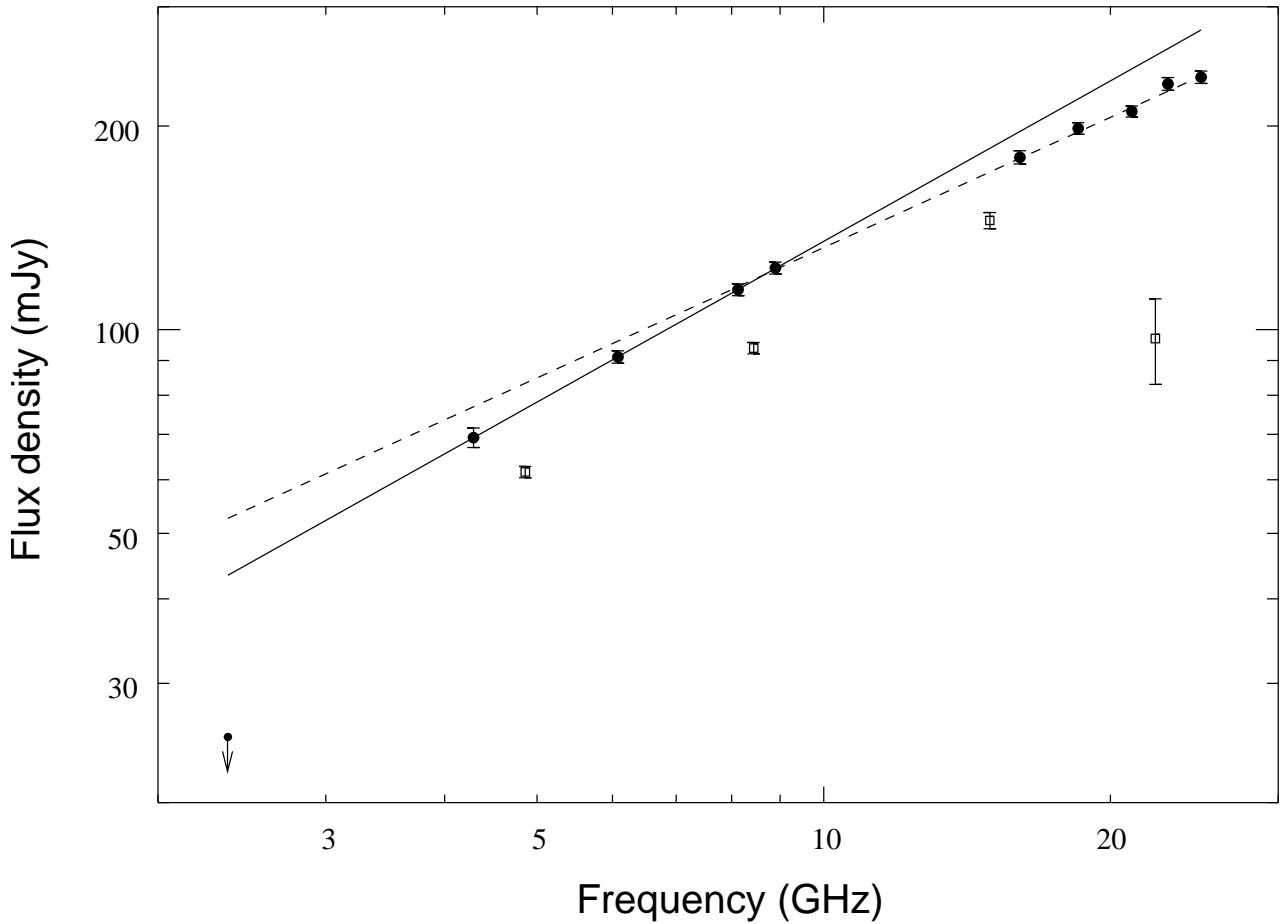}
\caption{\label{fig:spin} Flux density of IRAS 17347-3139 as a
  function of frequency. Filled circles represent ATCA data
  (2004 March). Open squares represent VLA data (1991 February,
  except the point at 22 GHz, taken on 2002 July). The
  solid line is the best fit for ATCA data at $\nu < 9$ GHz (spectral
  index = 0.79). The dashed line is the fit for $\nu > 16$ GHz
  (spectral index = 0.64)}
\end{figure*}

Initial calibration of the data was made using the MIRIAD package, 
while self-calibration (in both
phase and amplitude) and deconvolution was performed with Difmap and AIPS. 
After checking that the
resulting flux densities of IRAS 17347-3139 at each frequency were
compatible, within the errors, for data taken on March 4 and 10 (thus
ruling out
source variability at timescales of days
and significant calibration errors), we combined the data of
both periods to improve the uv coverage.
Table \ref{tab:fden} and Fig. \ref{fig:spin} summarize the final results of our measurements of flux density in
IRAS 17347-3139. 

\begin{table}
\caption{Continuum emission of IRAS 17347-3139}
\label{tab:fden}
\begin{tabular}{llll}
\hline
Frequency & Flux Density & Beam size & Beam P.A. \\
(GHz) & (mJy) & (arcsec) & (deg)\\
\hline
2.368 & $< 25$ & $16.6\times 4.8$ & $-33.1$\\ 
4.288 & $69.2\pm 2.3$ &  $10.0\times 2.7$ & $-29.0$\\ 
6.080 & $91.1\pm 1.9$ &  $7.1\times 2.0$ & $-29.4$\\
8.128 & $114.6\pm 2.3$ &  $5.7\times 1.3$ & $-25.7$\\
8.896 & $123.3\pm 2.5$ & $5.3\times 1.2$ & $-25.7$\\
16.064 & $180\pm 4$ & $2.2\times 0.6$ & $-15.3$\\
18.496 & $198\pm 4$& $2.0\times 0.5$ & $-16.2$\\
21.056 & $210\pm 4$ & $1.7\times 0.4$ & $-21.0$\\
22.976 & $231\pm 5$ & $1.5\times 0.4$ & $-25.6$\\
24.896 & $236\pm 5$ & $1.3\times 0.5$ & $-29.7$\\
\hline
\end{tabular}
\end{table}

The point at 2.4 GHz is given as an upper limit although the map
shows emission of $\simeq 10$ mJy near the position of the radio
continuum emission of IRAS 17347-3139 at other frequencies. However
there is a relatively strong extragalactic source, TXS 1734-314, within
the primary beam with a flux
density of $\simeq 995\pm 20$ mJy at 2.4 GHz and peak intensity 
located at $\alpha(2000)
=  17^h 37^m 43\fs 4$, $\delta(2000) = -31^\circ 31' 10''$
(but undetected at the other frequencies). 
TXS 1734-214 produces sidelobes stronger than  10 mJy,
and therefore, we cannot be certain that the emission apparently
associated to IRAS 17347-3139 at 2.4 GHz is real, and even if it is, we
cannot determine its flux density properly. 
Source confusion makes snapshot data from a linear array like ATCA
difficult to calibrate and process \citep{bur95}, especially at the
deconvolution stage. The upper limit we give in Table \ref{tab:fden}
is the flux density of the strongest sidelobe in the map.
A more accurate determination or upper limit on the 2.4~GHz flux density of
IRAS 17347-3139 would require imaging the field of
view with better uv coverage.
With this upper limit we estimate a spectral index $\alpha\geq 1.7$
between 2.4 
and 4.3 GHz which is consistent with the ionized
region being in the optically thick regime at 2.4 GHz, since
we expect a spectral index $\alpha\simeq 2$ in that case. 

For the detected emission, from 4.3 to 24.9 GHz, 
the spectrum of IRAS 17347-3139 rises with increasing frequency.
This is not 
consistent with our result in DG04 using the VLA 
in which the flux density seemed
to drop at 22 GHz. We think this discrepancy is most likely due to
calibration errors in the VLA data at these high frequencies and at
low elevation, or to extended emission ($\ga 5''$)
being resolved out by the VLA.

A careful analysis of the spectral index (Fig. \ref{fig:spin})
shows that it is decreasing
at the highest frequencies. We estimate an index of $0.79\pm 0.04$
between 4.3 and 8.9 GHz (consistent with our VLA estimate of
$0.76\pm 0.03$, DG04) whereas the values of flux densities at $\nu >
16$ GHz are slightly but significantly lower than those expected from
the extrapolation of the lower frequency fluxes. We derive a spectral
index of $0.64\pm 0.06$ between 16.1 and 24.9 GHz.

Therefore it seems that the radio continuum emission is becoming
optically thin around the highest frequencies we sampled, as suggested
in DG04. Of course an accurate determination of the turnover
frequency would require observations at frequencies even higher than
24 GHz to reach the region of the spectrum where the spectral index
would be $\simeq 0$.

It is noticeable that the spectral index does not change much
over a wide range of frequencies (from 4.3 to 24.9 GHz), 
rather than showing a rapid transition
between the optically thick and thin regimes, which would occur in the
case of an homogeneus H{\sc ii} region. Such relatively constant
spectral indices are expected in ionized regions with a power-law
decrease of electron density with distance from the central star
\citep{oln75,pan75,wri75}. In our case an average spectral index of
$\alpha\simeq 
0.7$ would require the electron density to vary as $n_e\propto
r^{-2.1}$, where $r$ is the distance to the central star. 
Similar spectral indices are also expected if the radiocontinuum emission
arises from a collimated wind, as suggested for the radio core of M2-9 
\citep{kwo85,lim03}. However, as discussed in DG04, in the case of 
IRAS 17347-3139 the derived mass
loss-rate under the assumption of a wind is 
$\dot{M}\simeq 1\times 10^{-4} (D/{\rm kpc})^{3/2}$ $M_\odot$
yr$^{-1}$, much higher than expected in central stars of PNe and proto-PNe ($\dot{M}\la 10^{-7}$ $M_\odot$;
\citealt{pat91,vas94}), and 
$\ga 2$ orders of magnitude larger than
the value derived for the core of M2-9 by \cite{lim03}, which
was already difficult to explain even if the main source 
of mass-loss
were an AGB star in a symbiotic system. 
Therefore, in principle we favor that the radio 
continuum emission arises from an 
ionized nebula, although higher-resolution observations 
would be useful to
determine whether there is a collimated radio outflow in IRAS
17347-3139.

From the decrease in the spectral index at $\simeq 20$
GHz we can estimate the turnover to be around this frequency.
As mentioned in DG04, assuming a turnover frequency of $\simeq 20$ GHz we
derive an emission measure $EM = 1.5\times 10^9$ cm$^{-6}$ pc.
In that paper we estimated a distance of 0.8 kpc based on two
methods (evolutionary arguments, and statistical
distance). Unfortunately, the statistical distance was incorrect, and
it should be $\simeq 6$ kpc instead of 0.8 kpc \citep[using the scale
of][based on the mass-radius relationship]{zha95}. 
Therefore we have to
consider a distance range of 0.8-6 kpc in the determination of other
parameters. We also note that statistical distances for PNe are very
uncertain and they can vary up to more than 
a factor of $\simeq 3$ depending on the 
particular scale used \citep[cf.][and references
therein]{ben01,phi02}. 
With a distance range of 0.8-6 kpc we derive 
 an average electron 
density $n_e = 4.1\times 10^5-1.1\times 10^6$
cm$^{-3}$ for a size of $0\farcs 3$.

\section{Discussion}

\label{sec:dis}
IRAS 17347-3139 is an unusual PN, since it is one of the only two
known to harbour
water maser emission. 
Therefore it is important to look for any other observational
characteristics that may be
related to the presence of water maser emission.

Several observational signatures, that we will discuss here,
strongly suggest that IRAS 17347-3139
is one of the youngest PNe known.
First, the turnover frequency of the radio continuum emission in 
IRAS 17347-3139, $\simeq 20$ GHz, is abnormally high as compared with
other PNe, in which the radio emission
is optically thin for frequencies $> 10$ GHz 
\citep[e.g.,][]{hug67,tay87,zij89,aaq91}. Noticeable
exceptions are AFGL 618 and Hb 12, which
have  turnover frequencies $\geq 22$ GHz \citep{kwo84} and $\simeq 30$
GHz \citep{aaq91}, respectively. These
objects have been proposed to be
very young PN \citep[$\simeq 100$ yr in the case of AFGL 618;][]{kwo84}.
For IRAS 17347-3139, the
high turnover frequency is also consistent with this source being a
very young PN, since the high emission measure that it implies is a
signature of youth \citep{kwo81}.

\begin{figure}
\includegraphics{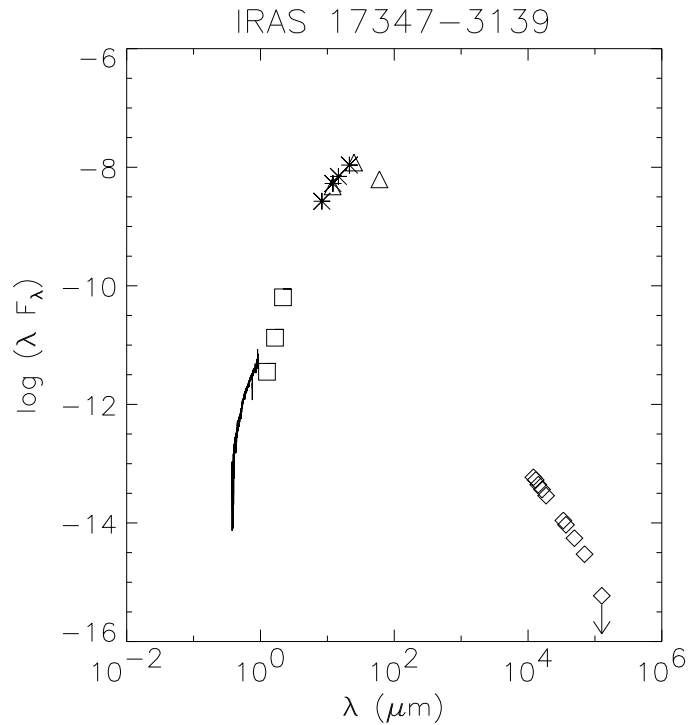}
\caption{\label{fig:sed} Spectral energy distribution of IRAS
  17347-3139, adapted from \citet{sua04}. The solid line is the
optical spectrum. The squares, asterisks and triangles represent the
2MASS, MSX, and IRAS data, respectively. The diamonds are the ATCA
radio continuum measurements
(this paper).}
\end{figure}

The optical spectrum of IRAS 17347-3139 shows very few lines. 
Of particular interest is the absence of H$\alpha$ emission 
in the spectrum shown in the atlas of \cite{sua04}, while all
other objects classified as PN and ``transition objects'' 
in this atlas show this line. Only [S
{\sc iii}] at $\lambda 9069$ \AA\ 
is clearly detected by \cite{sua04}.  The absence
of optical features in IRAS 17347-3139 can
be attributed to an extremely high extinction, which is
also evident in its spectral energy distribution (SED;
Fig. \ref{fig:sed}). 
The SED of IRAS 17347-3139 shows only a peak,
at $\simeq 25$ $\mu$m, 
with a sharp decrease toward optical wavelengths. The fact that this
object shows ionization while still keeping a very thick envelope to
produce such a strong extinction suggests a very rapid evolution and
therefore implies that the progenitor star was relatively massive 
\citep{blo95}. This is also supported by the bipolar morphology of the nebula
(DG04), since bipolar PNe seem to be associated with the evolution of
more massive stars than those producing elliptical or spherical PNe \citep{cor95}.

It is also interesting that the radio flux density of IRAS  
17347-3139 seems to have risen in the 13 years between the VLA and
ATCA observations (Fig. \ref{fig:spin}). Assuming that the opacity of the ionized
region has not varied significantly (the spectral index is similar in
both epochs), a possible explanation of this increase in flux
density could be the expansion of the  ionized region. 
At a frequency of $\simeq 8.4$ GHz,  
the increase is of a factor of $\simeq 1.26$, which 
would correspond to expansion by a factor of 1.12 in radius in
13 years. 
At this rate, and assuming a constant
expansion velocity, the kinematical age of the ionized region is only
$\simeq 120$ yr. The corresponding linear expansion velocity, assuming a
radius of $\simeq 0\farcs 15$, at a distance of $\simeq 0.8-6$ kpc would
be $\simeq 5-40$ km
s$^{-1}$ ($\simeq 1-8$ AU yr$^{-1}$). 
This estimate is  consistent with the expansion velocities 
found in PNe 
\citep[typically $\simeq 10-35$ km 
s$^{-1}$;][]{sab84,wei89}.
We also note that radio brightening has been
reported in AFGL 618 \citep{kwo81a}, also interpreted as expansion of
its associated ionized region. An increase of the flux density at radio
wavelengths is expected in some models of PN evolution \citep{vol85} during
the first few hundred years, especially for the most massive cases \citep[see
Fig. 14 in][]{vol85} while the ionized nebula expands but the
optical depth is still high.

The evidence we have presented above, indicating that IRAS 17347-3139
is an extremely young PN,
is consistent with the presence of water maser emission since 
water molecules are
thought to be destroyed in $\simeq 100$ yr \citep{gom90}. Given the
characteristics found in IRAS 17347-3139, we have checked for the 
possible presence
of water masers in AFGL 618 and Hb 12, to see if a rising radio
spectrum at 
$\simeq 20$ GHz could be an indication of a higher probability of maser
emission. 
Since we did not
find in the literature any report of water maser observations toward
Hb 12, and given that the $3\sigma$ uppper limit of $\geq 2.1$ Jy 
obtained by \cite{wou93} towards AFGL 618
(IRAS 04395+3601) is relatively high, 
we took spectra of the 6$_{16} \rightarrow$5$_{23}$
transition of the water molecule  
(rest frequency 22235.080 MHz), with NASA's 70m DSN antenna in Robledo
de Chavela (Spain), on 2005 May 20, toward these two PNe. 
These spectra show no emission,
with 3$\sigma$ upper limits of 160 mJy between $V_{\rm LSR}=$ -128.9 and
+86.9 km s$^{-1}$ for AFGL 618, and 130 mJy between $V_{\rm LSR}=$ -112.7 and
+103.0 km s$^{-1}$ for Hb 12 (with a velocity resolution of 0.56 km
s$^{-1}$ in both cases).

We note that K3-35, 
the other PN known to show
water maser emission
\citep{mir01},
does not have a rising spectrum at this
frequency \citep[its turnover frequency is $\simeq 10$ GHz;][]{aaq91}
nor a 
strongly obscured optical spectrum 
\citep[it shows emission of
H$\alpha$ and other prominent optical lines;][]{mir00}, which
could indicate that this source may be in a later evolutionary stage than 
IRAS 17347-3139. 

All the unusual characteristics of IRAS 17347-3139 suggest that this
could be one of the youngest ($< 100$ year) PNe known, and therefore,
it is one of the most appropriate objects to further study the early
stages of this phase of 
stellar evolution, together with other PNe with rising spectra
at high radio frequencies.

\section*{Acknowledgments}

GA, IdG, JFG, LFM, and JMT are partially supported by grant AYA2002-00376
of the Spanish MEC (co-funded by FEDER funds). OS is partially
supported by grant AYA2003-09499 of MEC. GA and LFM
are also supported by Junta de Andaluc\'{\i}a. YG acknowledges the
support from DGAPA, UNAM, and CONACyT, Mexico. IdG acknowledges the
support of a Calvo Rod\'es 
Fellowship from the Instituto Nacional de T\'ecnica Aeroespacial. 
This paper is partly based on observations taken during
``host-country'' allocated time at Robledo de Chavela; this time is managed
by the LAEFF of INTA, under agreement with NASA/INSA.


\begin{thebibliography}{}
\bibitem[Aaquist \& Kwok (1991)]{aaq91} Aaquist O. B.,  Kwok
  S., 1991, ApJ, 378, 599
\bibitem[Bensby \& Lundstr\"om (2001)]{ben01} Bensby T., 
  Lundstr\"om I., 2001, A\&A, 374, 599
\bibitem[Bl\"oker (1995)]{blo95} Bl\"ocker T., 1995, A\&A, 299, 755
\bibitem[Burgess \& Hunstead (1995)]{bur95} Burgess A. M., Hunstead
  R. W., 1995, Pub. Astron. Soc. Aust., 12, 227
\bibitem[Corradi \& Schwarz (1995)]{cor95} Corradi R. L. M., 
  Schwarz H. E., 1995, A\&A, 293, 871
\bibitem[de Gregorio-Monsalvo et al. (2004)]{dg04} de
    Gregorio-Monsalvo I., G\'omez Y., Anglada G., Cesaroni R.,
    Miranda L. F.,  G\'omez J. F., Torrelles J. M., 2004, ApJ, 601,
    921 (DG04)
\bibitem[G\'omez et al. (1990)] {gom90}  G\'omez Y., Moran J. M.,  
  Rodr\'{\i}guez L. F., 1990, Rev. Mexicana Astron. Astrofis., 20, 55. 
\bibitem[Hasegawa et al. (2000)]{Has00} Hasegawa T., Volk K.,  
  Kwok S., 2000, ApJ, 532, 994
\bibitem[Hughes (1967)]{hug67} Hughes M. P., 1967, ApJ, 149, 377 
\bibitem[Kwok \& Bignell (1984)]{kwo84} Kwok S.,  Bignell
  R. C., 1984, ApJ, 276, 544
\bibitem[Kwok \& Feldman (1981)]{kwo81a} Kwok S.,  Feldman
  P. A., 1981, ApJ, 247, L67
\bibitem[Kwok, Purton, \&  Keenan (1981)]{kwo81} Kwok S., Purton
  C. R., Keenan D. W., 1981, ApJ, 250, 232
\bibitem[Kwok et al. (1985)]{kwo85} Kwok S., Purton C. R., Matthews
  H. E., Spoelstra T. A. T., 1985, A\&A, 144, 321
\bibitem[Lane et al. (1987)]{lan87} Lane A. P., Johnston K. J.,
  Bowers 
  P. F., Spencer J. H.,  Diamond P. J., 1987, ApJ, 323, 756
\bibitem[Lewis (1989)]{lew89} Lewis B. M., 1989, ApJ, 338, 234
\bibitem[Lim \& Kwok (2003)]{lim03} Lim J., Kwok S., 2003, in
  Corradi R. L. M.,
  Mikolajewska R., Mahoney T. J., eds, ASP Conf. Ser. Vol. 303,
  Symbiotic Stars Probing Stellar 
  Evolution. Astron. Soc. Pac., San
  Francisco, p. 437 
\bibitem[Miranda et al. (2000)] {mir00} Miranda L. F., Fern\'andez
  M., Alcal\'a J. M., Guerrero M. A., Anglada G., G\'omez Y.,
  Torrelles J.  M., Aaquist O. B., 2000, MNRAS, 311, 748
\bibitem[Miranda et al. (2001)] {mir01} Miranda L.F., G\' omez Y.,  
  Anglada G.,  Torrelles J. M., 2001, Nature, 414, 284  
\bibitem[Olnon (1975)]{oln75} Olnon F. M., 1975, A\&A, 39, 217
\bibitem[Panagia \& Felli (1975)]{pan75} Panagia N., Felli M., 1975,
  A\&A, 39, 1
\bibitem[Patriarchi \& Perinotto (1991)]{pat91} Patriarchi P.,
  Perinotto M., 1991, A\&AS, 91, 325 
\bibitem[Phillips (2002)]{phi02} Phillips J. P., 2002, ApJS, 139, 199
\bibitem[Reid \& Moran (1981)]{rei81} Reid M. J., Moran
  J. M., 1981, ARA\&A, 19, 231
\bibitem[Sabaddin (1984)]{sab84} Sabaddin F., 1984, A\&AS, 58, 273
\bibitem[Su\'arez et al. (2005)]{sua04} Su\'arez O., Garc\'{\i}a-Lario
  P., Manchado  A., Manteiga  M., Ulla A.,
Pottasch S. R., 2005, A\&A,
  submitted
\bibitem[Taylor, Pottasch, \&  Zhang (1987)]{tay87} Taylor A. R.,
  Pottasch S. R., Zhang C. Y., 1987, A\&A, 171, 178
\bibitem[Vassiliadis \& Wood (1994)]{vas94} Vassiliadis E., Wood
  P. R., 1994, ApJS, 92, 125 
\bibitem[Volk \& Kwok (1985)]{vol85} Volk K.,  Kwok
  S., 1985, A\&A, 153, 79
\bibitem[Weinberger (1989)]{wei89} Weinberger R., 1989,  A\&AS, 78,
  301
\bibitem[Wouterloot, Brand, \& Fiegle (1993)]{wou93} Wouterloot
  J. G. A., Brand J., Fiegle K., 1993, A\&AS, 98, 589
\bibitem[Wright \& Barlow (1975)]{wri75} Wright A. E., Barlow
  M. J., 1975, MNRAS, 170, 41
\bibitem[Zhang (1995)]{zha95} Zhang C. Y., 1995, ApJS, 98, 659
\bibitem[Zijlstra, Pottasch, \& Bignell (1989)]{zij89} Zijlstra
    A. A., Pottasch S. R., Bignell C., 1989, A\&AS, 79, 329

\end{thebibliography}
\end{document}